\documentclass[twocolumn]{aastex631}

\shorttitle{Surrounded by Giants}
\shortauthors{Stephen R. Kane}

\begin{document}

\title{Surrounded by Giants: Habitable Zone Stability Within the HD
  141399 System}

\author[0000-0002-7084-0529]{Stephen R. Kane}
\affiliation{Department of Earth and Planetary Sciences, University of
  California, Riverside, CA 92521, USA}
\email{skane@ucr.edu}

%%%%%%%%%%%%%%%%%%%%%%%%%%%%%%%%%%%%%%%%%%%%%%%%%%%%%%%%%%%%%%%%%%%%

\begin{abstract}

The search for exoplanets has revealed a diversity of planetary system
architectures, the vast majority of which diverge significantly from
the template of the solar system. In particular, giant planets beyond
the snow line are relatively rare, especially for low-mass stars,
placing the solar system within a small category of systems with
multiple giant planets at large separations. An exoplanetary system of
note is that of HD~141399, consisting of a K dwarf host star that
harbors four giant planets with separations extending to
$\sim$4.5~AU. The architecture of the system creates a complex pattern
of mean motion resonances and gravitationally perturbed regions that
may exclude the presence of other planets, including within the
Habitable Zone of the system. Here, we present the results of
dynamical simulations that explore the interaction of the known
planets of the system, their apsidal trajectories, resonance
locations, and dynamical evolution. We further investigate the results
of injecting Earth-mass planets and provide the regions of dynamical
viability within the Habitable Zone where terrestrial planets may
maintain long-term stability. We discuss these results in the context
of the importance of giant planets for volatile delivery and
planetary habitability considerations.

\end{abstract}

\keywords{astrobiology -- planetary systems -- planets and satellites:
  dynamical evolution and stability -- stars: individual (HD~141399)}

%%%%%%%%%%%%%%%%%%%%%%%%%%%%%%%%%%%%%%%%%%%%%%%%%%%%%%%%%%%%%%%%%%%%

\section{Introduction}
\label{intro}

The origin and evolution of planetary architectures are key areas of
study within exoplanetary science. The large number of discovered
planets allows a statistical analysis of these architectures and a
direct comparison with the solar system
\citep{ford2014,winn2015,horner2020b,kane2021d,mishra2023a,mishra2023b}. In
particular, we are beginning to understand the prevalence of Jupiter
analogs
\citep{wittenmyer2011a,wittenmyer2020b,fulton2021,rosenthal2021},
largely from the long baseline of radial velocity (RV) observations
\citep{kane2007a,ford2008a,wittenmyer2013a}. For orbits of 1--10~AU,
RV surveys show that giant planets are found orbiting about 6\% of
solar-type stars \citep{wittenmyer2016c}. Combined analysis of RV
surveys, microlensing, and direct imaging surveys also shows that
giant planet frequencies are likely in the 10\% range for 1--100~AU
\citep{clanton2016}. Other combined analyses shows that the giant
planet occurrence rate peaks at orbital separations of 1--3~AU
\citep{fernandes2019a}. Direct imaging surveys are able to probe
orbits in the range 10--100~AU, finding giant planets are more common
around more massive host stars \citep{nielsen2019c}. The occurrence of
giant planets is also correlated with other system properties, such as
the metallicity of the host star
\citep{fischer2005b,johnson2010d,buchhave2018} and the the presence of
super-Earths \citep{bryan2019}. However, there has been an observed
dearth of Jupiter analogs around M dwarf stars
\citep{endl2006b,pass2023b}, although exceptional cases have been
detected \citep{endl2022}.  These insights into the occurrence of
Jupiter analogs are critical for placing our solar system in a broader
context of planetary architectures \citep{levison1998}.

Planetary habitability can depend on the entire system architecture,
not only in the context of volatile delivery, but also in terms of the
dynamical interactions (past and present) between a planet that is a
promising candidate for habitability and any other planets orbiting
the same star. For example, the presence of Jupiter analogs can limit
the viability of terrestrial planet occupation with the Habitable Zone
(HZ) of the system
\citep{kopparapu2010,obertas2017,georgakarakos2018,hill2018,agnew2019,kane2019e,kane2020b}. The
distribution of planetary architectures speaks to critical issues
within the fields of planetary science and astrobiology, such as
determining if the architecture of our solar system is typical, and
the subsequent implications for volatile delivery and planetary
habitability. Systems that harbor multiple giant planets are therefore
of particular interest in how they may facilitate or truncate
planetary habitability within the system.

Given the importance, and relative scarcity, of giant planets, a
planetary system of particular interest is that of HD~141399. The host
star is an early K dwarf and was found to harbor four giant planets
\citep{vogt2014b} with masses in the range $\sim$0.45--1.36~$M_J$. The
planets were detected from RV observations of the host star, and so
the measured planetary masses are lower limits. The system properties
were further refined by \citet{hebrard2016b} and
\citet{rosenthal2021}, whose increased observational baseline and
improved RV precision produced a robust orbit for the outermost
planet, with a period of almost 10 years. Such a large family of giant
planets is exceptionally rare, providing a useful opportunity to study
the dynamics of interacting giant planets, that can often lead to
scattering events and eccentric orbits
\citep{ford2008c,carrera2019b}. The system also allows an examination
of the extreme end of how giant planets and their associated orbital
resonances may influence the orbital evolution of terrestrial planets.

In this paper, we present the results of an extensive dynamical
analysis of the HD~141399 planetary system, including their orbital
evolution and effect on possible terrestrial planets within the HZ.
Section~\ref{arch} describes the architecture and HZ of the system, as
well as the locations of mean motion resonance
(MMR). Section~\ref{stab} presents the results of the dynamical
simulations, both intrinsically for the known system architecture, and
the influence of that architecture for an Earth-mass planet within the
HZ. The potential further work and implications of these results for
planetary habitability are discussed in Section~\ref{discussion}, and
we provide concluding remarks in Section~\ref{conclusions}.

%%%%%%%%%%%%%%%%%%%%%%%%%%%%%%%%%%%%%%%%%%%%%%%%%%%%%%%%%%%%%%%%%%%%

\section{System Architecture}
\label{arch}

Here we describe the HD~141399 system architecture, calculate the HZ,
and provide the locations of MMR for the planets.

%%%%%%%%%%%%%%%%%%%%%%%%%%%%%%%%%%%%%%%%%%%%%%%%%%%%%%%%%%%%%%%%%%%%

\subsection{Orbits and Habitable Zone}
\label{orbits}

The central body of the HD~141399 system consists of a K0V star
\citep{vogt2014b,hebrard2016b}. We adopt the stellar mass of $M_\star
= 1.091$~$M_\odot$ and the planetary properties (masses and orbits)
from \citet{rosenthal2021}. The masses and orbital properties of the
planets are summarized in Table~\ref{tab:planets}, which shows that
all of the planets are more massive than Saturn and have orbits that
are close to circular. Note that, since these planetary masses are
derived from RV observations, these are lower limits on their true
masses. Both \citet{vogt2014b} and \citet{hebrard2016b} conducted
stability tests by assuming coplanarity of the orbits and decreasing
the inclination of the system with respect to the plane of the
sky. Their results indicate that the inclination of the system is
constrained to be larger than $\sim$10$^\circ$, and so the true
inclination of the planetary orbits remain largely unknown.

\begin{deluxetable}{cccccc}
  \tablecolumns{6}
  \tablewidth{0pc}
  \tablecaption{\label{tab:planets} HD~141399 planetary properties.}
  \tablehead{
    \colhead{Planet} &
    \colhead{$P$} &
    \colhead{$a$} &
    \colhead{$e$} &
    \colhead{$\omega$} &
    \colhead{$M_p \sin i$} \\
    \colhead{} &
    \colhead{(days)} &
    \colhead{(AU)} &
    \colhead{} &
    \colhead{($^\circ$)} &
    \colhead{($M_J$)}
  }
  \startdata
  b &   94.375 & 0.4176 & 0.053  &   8.0 & 0.452 \\
  c &  201.776 & 0.693  & 0.0465 &  25.0 & 1.329 \\
  d & 1074.8   & 2.114  & 0.044  & 297.0 & 1.263 \\
  e & 3336     & 4.50   & 0.047  &   0.0 & 0.644
  \enddata
  \tablecomments{Planetary properties extracted from
    \citet{rosenthal2021}.}
\end{deluxetable}

\begin{figure}
  \includegraphics[angle=270,width=8.5cm]{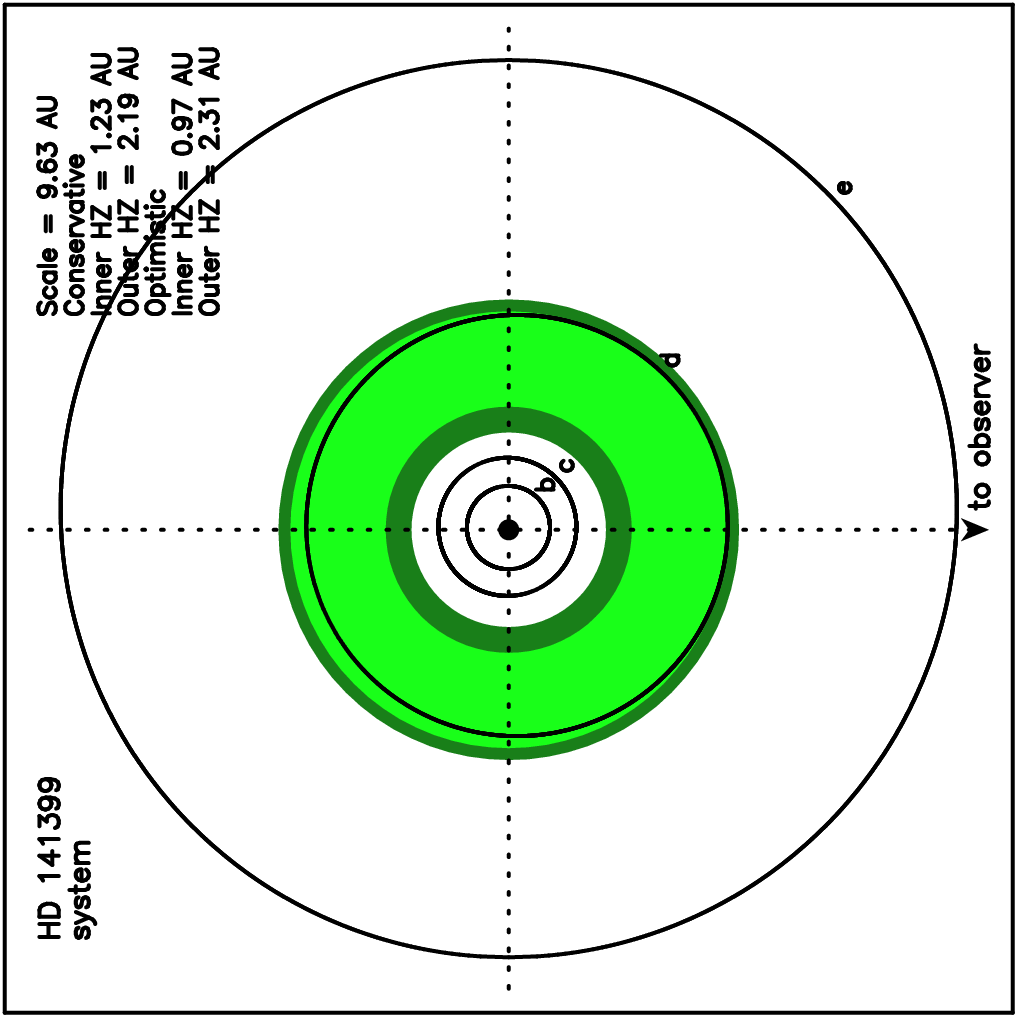}
  \caption{HZ and planetary orbits in the HD~141399 system, where the
    orbits are labeled by planet designation. The extent of the HZ is
    shown in green, where light green and dark green indicate the CHZ
    and OHZ, respectively. The scale of the figure is 9.63~AU along
    each side.}
  \label{fig:hz}
\end{figure}

\begin{deluxetable*}{llrrrrrrrrrrrr}
  \tablecolumns{14}
  \tablewidth{0pc}
  \tablecaption{\label{tab:mmr} Mean motion resonance locations for
    the HD~141399 planets.}
  \tablehead{
    \colhead{Planet} &
    \colhead{$a$ (AU)} &
    \colhead{3:1} &
    \colhead{5:2} &
    \colhead{7:3} &
    \colhead{2:1} &
    \colhead{3:2} &
    \colhead{7:5} &
    \colhead{5:7} &
    \colhead{2:3} &
    \colhead{1:2} &
    \colhead{3:7} &
    \colhead{2:5} &
    \colhead{1:3}
  }
  \startdata
  b & 0.4176 & 0.201 &  0.227 &  0.237 &  0.263 &  0.319 &  0.334 &  0.523 &  0.547 &  0.663 &  0.735 &  0.769 &  0.869 \\
  c & 0.693  & 0.333 &  0.376 &  0.394 &  0.437 &  0.529 &  0.554 &  0.867 &  0.908 &  1.100 &  1.219 &  1.277 &  1.441 \\
  d & 2.114  & 1.016 &  1.148 &  1.202 &  1.332 &  1.613 &  1.689 &  2.646 &  2.770 &  3.356 &  3.719 &  3.894 &  4.397 \\
  e & 4.50   & 2.163 &  2.443 &  2.558 &  2.835 &  3.434 &  3.596 &  5.632 &  5.897 &  7.143 &  7.916 &  8.289 &  9.360 \\
  \enddata
\end{deluxetable*}

We also calculate the HZ of the system, based upon an effective
temperature of $T_\mathrm{eff} = 5542$~K and stellar luminosity of
$L_\star = 1.637$~$L_\odot$, extracted from Gaia data release 2
\citep{brown2018}. The HZ may be divided into the conservative HZ
(CHZ) and optimistic HZ (OHZ), the latter of which is based upon
assumptions regarding the prevalence of surface liquid water for Venus
and Mars
\citep{kasting1993a,kane2012a,kopparapu2013a,kopparapu2014,kane2016c,hill2018,hill2023}. For
HD~141399, the HZ regions span the ranges 1.233--2.190~AU and
0.974--2.309~AU for the CHZ and OHZ, respectively. The extent of the
CHZ (light green) and OHZ (dark green) are shown in
Figure~\ref{fig:hz}, along with the orbits of the known planets.  The
outermost planet, planet e, lies beyond the snow line, estimated to be
$\sim$3.2~AU \citep{ida2005,kane2011d}. Planets c and d, the most
massive of the four planets, lie at either end of the HZ. In fact, the
orbit of planet d is slightly eccentric such that it crosses between
the outer OHZ and the outer edge of the CHZ. As such, the HZ of the
system is bookended by the gravitational influence of these two giant
planetary sentinels.

%%%%%%%%%%%%%%%%%%%%%%%%%%%%%%%%%%%%%%%%%%%%%%%%%%%%%%%%%%%%%%%%%%%%

\subsection{Locations of Mean Motion Resonance}
\label{mmr}

The periodic nature of gravitational perturbations that occur at
locations of MMR result in their having special significance within
complex system architectures \citep{petrovich2013,hadden2019b}. Such
locations have been the source of numerous investigations for the
solar system \citep{peale1976a} and exoplanetary systems
\citep{beauge2003b,goldreich2014}. MMR configurations can result from
numerous formation and dynamical processes \citep{batygin2015d}, and
may even be caused by planet-planet scattering events
\citep{raymond2008b}. More particularly, MMR locations can be the
source of secular perturbations that may lead to dynamical instability
for the associated planets \citep{matsumoto2012,batygin2013a}.

The relative even spacing of the HD~141399 planetary orbits, combined
with the high mass of the planets, produces a web of strong low order
MMR locations that extend out to $\sim$10~AU from the host star. These
MMR locations are shown in Table~\ref{tab:mmr} for all four of the
known planets. Of particular note are those MMR locations that align
closely between different planets, such as the 1:3 MMR for planet b
(0.869~AU) and the 5:7 MMR for planet c (0.867~AU). Also of note are
the close alignments of orbital semi-major axes with MMR locations,
such as the semi-major axis of planet c (0.693~AU) with the 1:2 MMR
for planet b (0.663~AU). The MMR locations within the system will play
an important role in the interpretation of the dynamical analyses for
the system.

%%%%%%%%%%%%%%%%%%%%%%%%%%%%%%%%%%%%%%%%%%%%%%%%%%%%%%%%%%%%%%%%%%%%

\section{Dynamical Stability}
\label{stab}

Here we describe the methodology of our dynamical simulations and
present the results for the known planets and an injected Earth-mass
planet within the HZ.

%%%%%%%%%%%%%%%%%%%%%%%%%%%%%%%%%%%%%%%%%%%%%%%%%%%%%%%%%%%%%%%%%%%%

\subsection{Dynamical Simulation Methodology}
\label{methods}

The dynamical simulations conducted for our work utilized the Mercury
Integrator Package \citep{chambers1999}, adopting the hybrid
symplectic/Bulirsch-Stoer integrator with a Jacobi coordinate system,
providing more accurate results for multi-planet systems
\citep{wisdom1991,wisdom2006b}. The time step for the integrations was
set to 0.1~days to ensure adequate resolution of perturbations
resulting from the innermost planet of the system. We assumed that the
orbits within the system are coplanar, and that the inclination of the
orbits are such that the measured minimum planetary masses are
reasonable approximations of their true masses.

The simulations were conducted in two primary configurations. First,
we conducted a suite of simulations for the known planets to evaluate
their long-term dynamical stability for a period of $10^8$ years. The
simulations explored a range of initial orbital configurations that
test semi-major axis values within the $1\sigma$ uncertainties of the
values provided by \citet{rosenthal2021}. Second, we conducted an
extensive suite of N-body integrations that explore the dynamical
viability of an injected Earth-mass planet, similar to the methodology
described by \citet{kane2019c,kane2021a}. Each of these integrations
were run for $10^7$ simulation years, significantly extending the
baseline of the integrations conducted by \citet{hebrard2016b}. The
injected planet was placed in a circular orbit using a range of
initial mean anomalies, and with semi-major axis values in the range
0.5--3.0~AU, in steps of 0.005~AU. This range of semi-major axes fully
encompasses the HZ, described in Section~\ref{orbits}. This grid of
initial conditions resulted in several thousand simulations to explore
orbital stability within the system HZ. The stability of the injected
planet was assessed through the planet surviving the full $10^7$~year
integration, where non-survival means that the injected planet was
captured by the gravitational well of the host star or ejected from
the system.

%%%%%%%%%%%%%%%%%%%%%%%%%%%%%%%%%%%%%%%%%%%%%%%%%%%%%%%%%%%%%%%%%%%%

\subsection{Stability of the Known Planets}
\label{known}

\begin{figure*}
  \begin{center}
    \includegraphics[angle=270,width=16.0cm]{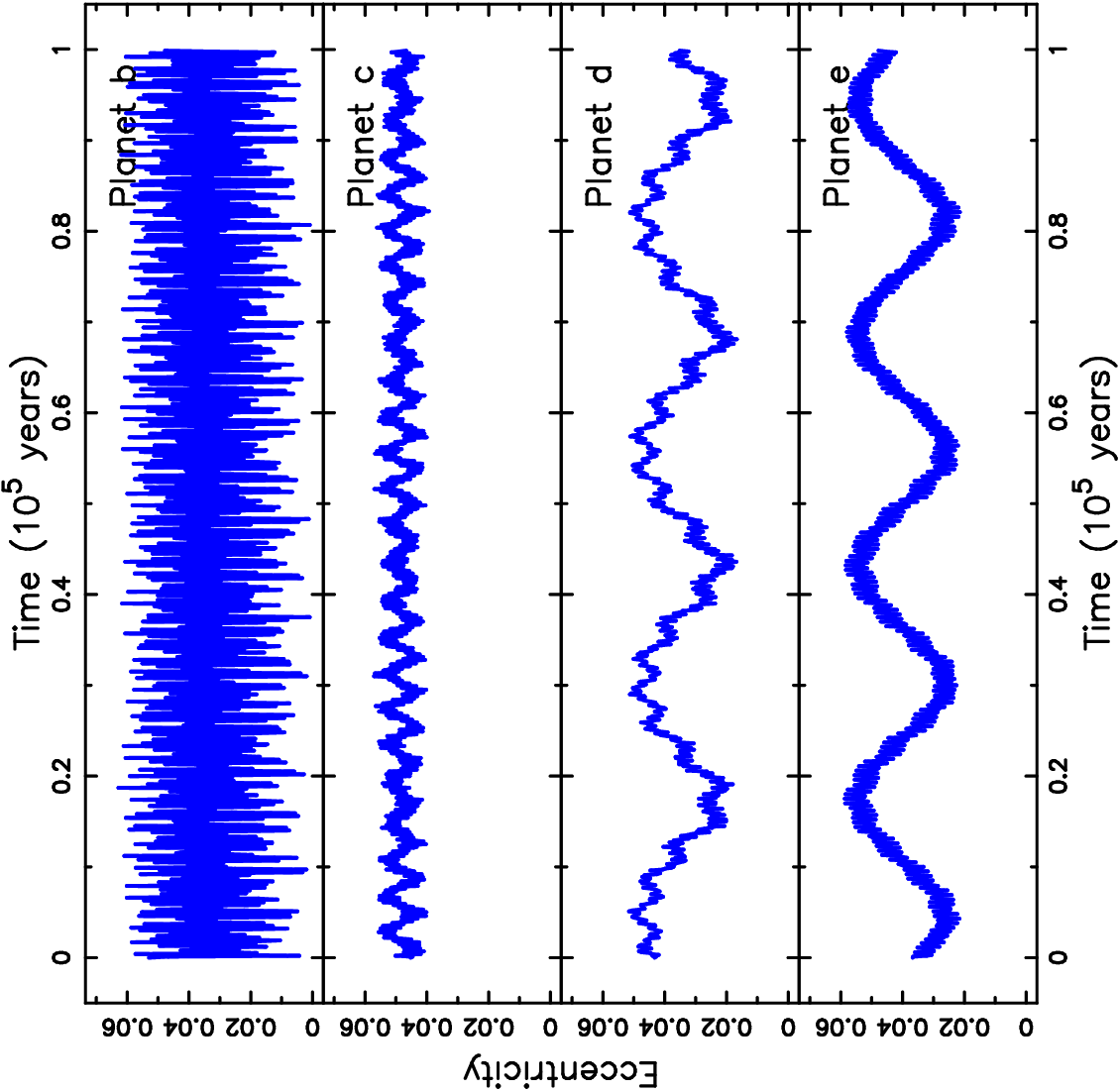}
  \end{center}
  \caption{Eccentricity evolution for the four known planets of the
    HD~141399 system. The data shown in each panel are the first
    $10^5$ years from a $10^8$ year integration.}
  \label{fig:known}
\end{figure*}

Given that the system consists of four giant planets, it is important
to consider angular momentum transfer that is occurring within the
system, and the timescale of such interactions
\citep{ford2001a,kane2014b}. Figure~\ref{fig:known} shows the first
$10^5$~years of a $10^8$~year simulation, indicating the nature of the
gravitational perturbations within the system. The planetary orbits
retain remarkable stability over the full $10^8$~year simulation, with
eccentricities for all planets remaining below 0.06. The eccentricity
evolution for each of the planets reveal periodic behavior over
various timescales. \citet{vogt2014b} noted an angular momentum
exchange between planets b and c with a period of $\sim$250 years,
based on a dynamical analysis of their best-fit Keplerian model at
that time. To ascertain the full periodic dynamical behavior within
the system, we conducted a Fourier analysis of the eccentricity
evolution for each planet. We find a chain of angular momentum
exchanges, starting with planets b and c whose eccentricity variations
have a period of $\sim$327 years. Planets c and d have eccentricity
variations of $\sim$4070~years, and planets d and e engage in
eccentricity variations with a period of $\sim$25590~years. Some of
these periodic variations may be interpreted within the context of MMR
locations (Table~\ref{tab:mmr}) such as the near 2:1 MMR between
planets b and c.

\begin{figure*}
    \begin{center}
        \begin{tabular}{cc}
            \includegraphics[angle=270,width=8.0cm]{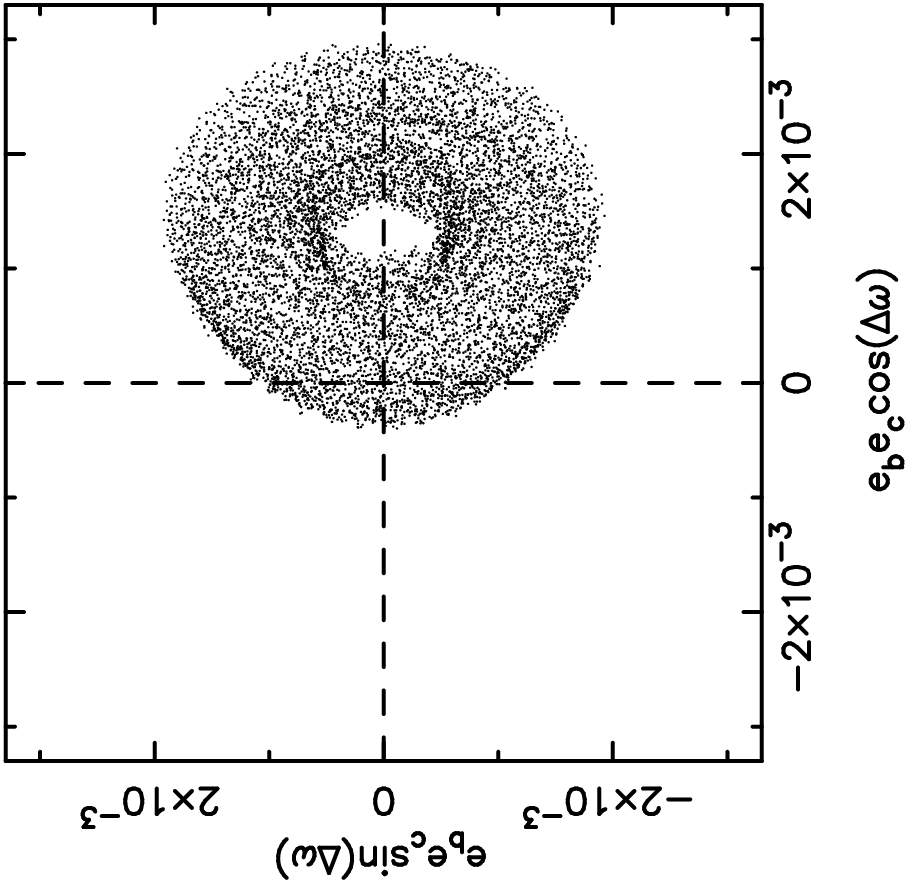} &
            \includegraphics[angle=270,width=8.0cm]{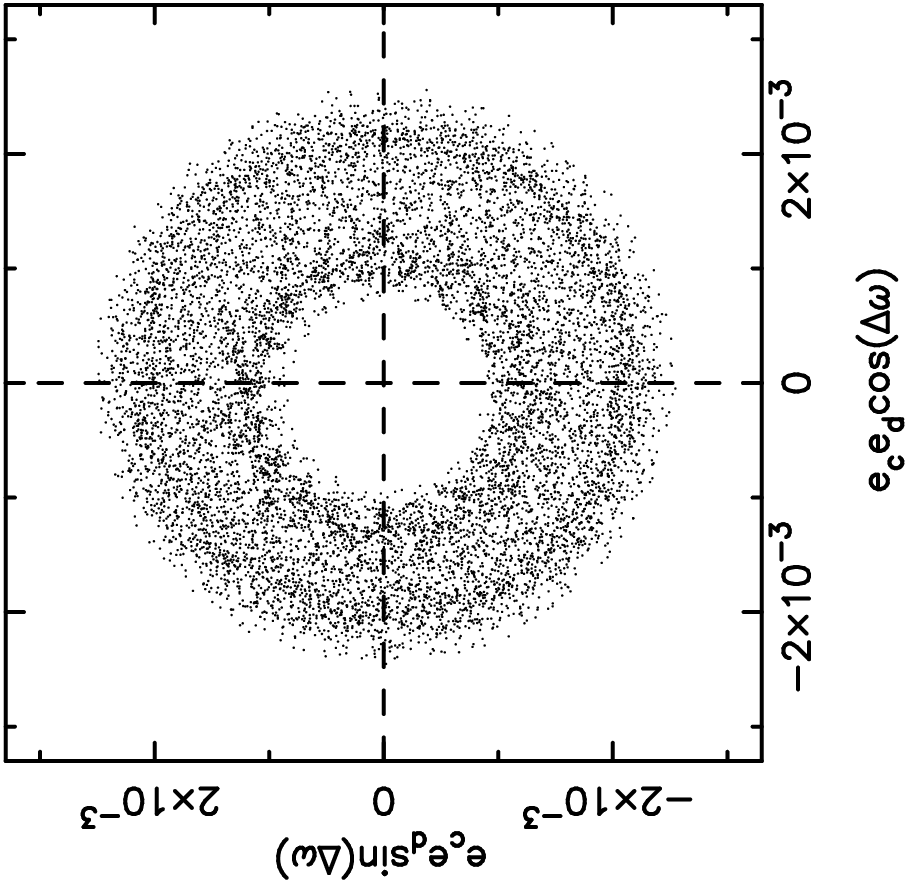} \\
            \includegraphics[angle=270,width=8.0cm]{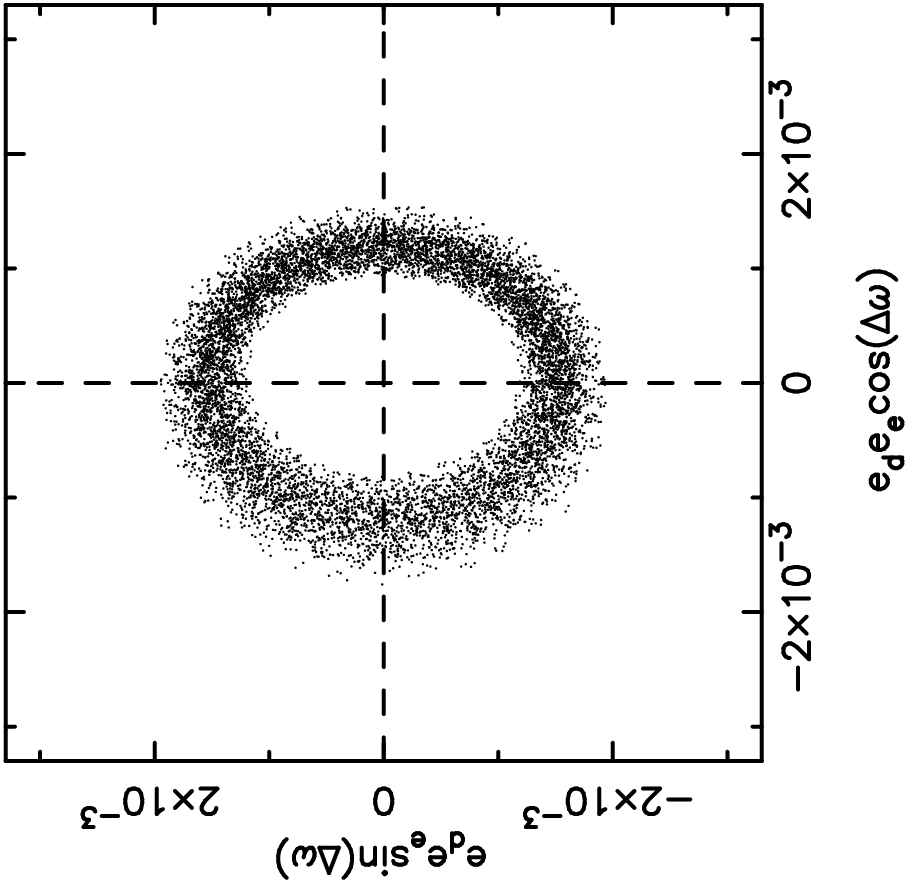} &
            \includegraphics[angle=270,width=8.0cm]{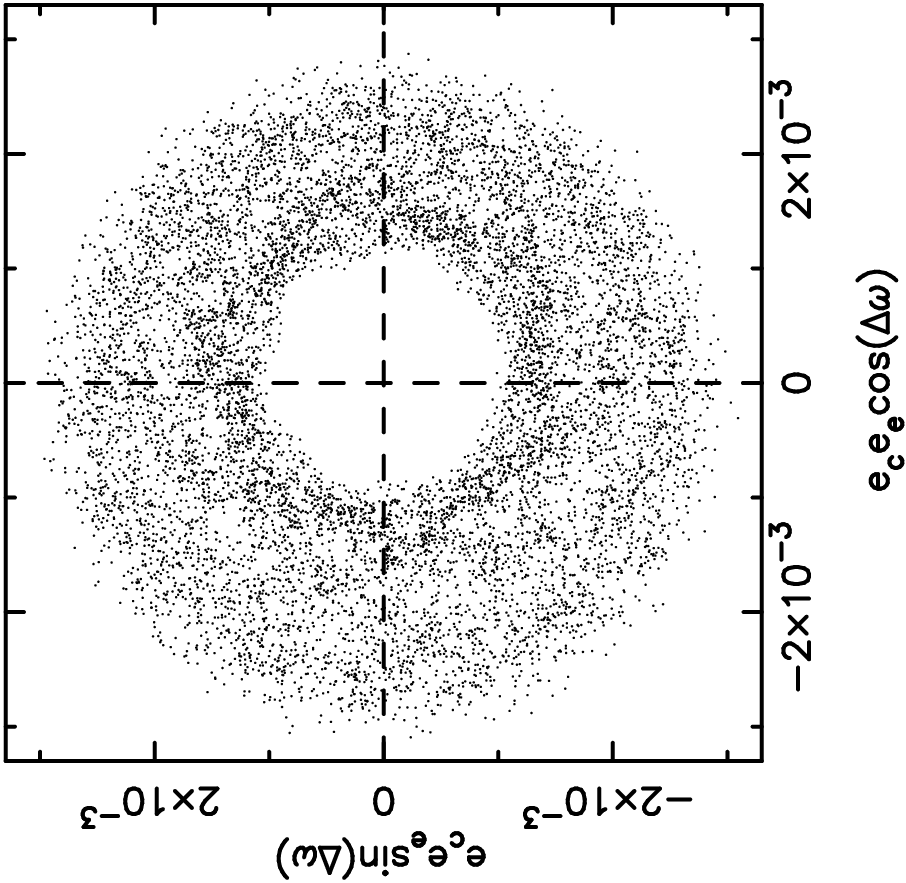}
        \end{tabular}
    \end{center}
  \caption{A polar plot of the apsidal trajectory for four planet
    pairs: b--c (top-left), c--d (top-right), d--e (bottom-left), and
    c--e (bottom-right). The data in each panel represent the first
    $10^6$~years of the simulation. The figure shows that the apsidal
    modes are librating for the b--c planet pair, and are circulating
    for all other planet pairs.}
  \label{fig:epsilon}
\end{figure*}

To investigate the dynamical behavior of the system further, we
calculated the trajectory of the apsidal modes for four planet pairs:
b--c, c--d, d--e, and c--e. Apsidal motion in the context of
interacting exoplanetary systems are described by
\citet{barnes2006a,barnes2006c}, where the two basic types of apsidal
behavior, libration and circulation, are separated by a boundary
called a secular separatrix \citep{barnes2006c,kane2014b}. The apsidal
trajectories for the chosen planet pairs during the initial $10^6$
simulation years are represented via their polar form in
Figure~\ref{fig:epsilon}. The top-right, bottom-left, and bottom-right
panels show clear indication of circulating apsidal motion for those
planet pairs, since the polar trajectories consistently encompass the
origin, and with no indication that they are near the separatrix
transition to libration. However, the apsidal trajectories for the
b--c planet pair, depicted in the top-left panel, show that they are
librating, which indicates that planets b and c are indeed in or near
the 2:1 MMR. Although the plots shown in Figure~\ref{fig:epsilon} only
represent the first $10^6$~years of the full $10^8$~year simulation,
we examined the evolution of both the eccentricity variations and
apsidal trajectories for the remainder of the simulation data and
found no significant changes in dynamical behavior.

%%%%%%%%%%%%%%%%%%%%%%%%%%%%%%%%%%%%%%%%%%%%%%%%%%%%%%%%%%%%%%%%%%%%

\subsection{Stability Within the Habitable Zone}
\label{hz}

\begin{figure*}
  \begin{center}
    \includegraphics[angle=270,width=16.0cm]{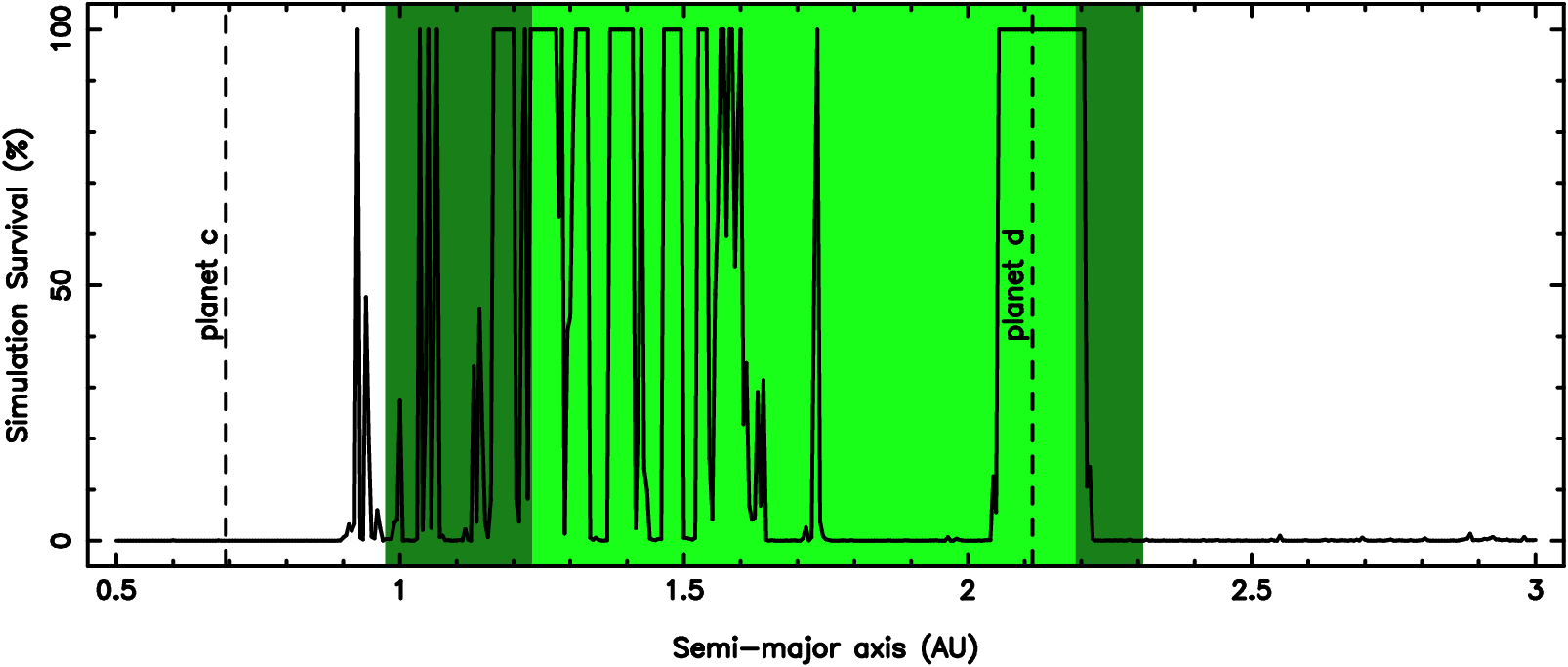} \\
    \includegraphics[angle=270,width=16.0cm]{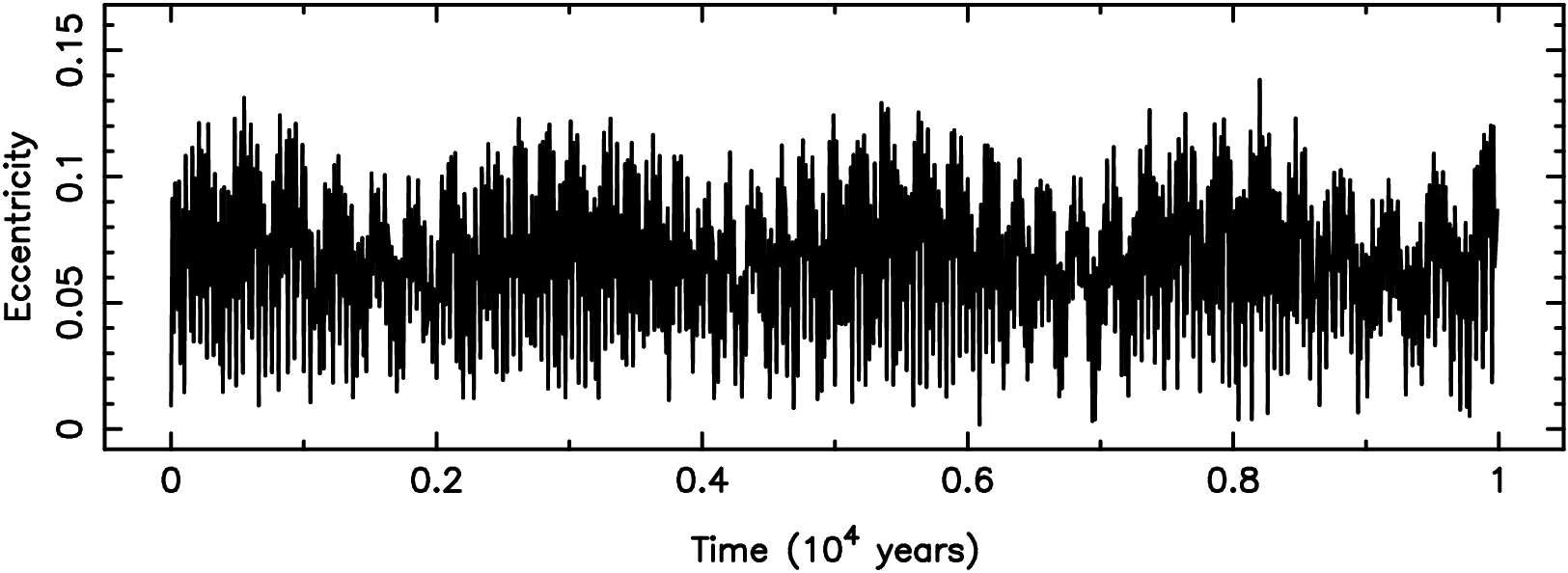}
  \end{center}
  \caption{Top: Percentage of the simulation that the injected
    Earth-mass planet survived as a function of semi-major axis, shown
    as a solid line. As for Figure~\ref{fig:hz}, the CHZ is shown in
    light green and the OHZ is shown in dark green. The vertical
    dashed lines indicate the semi-major axes for planets c and
    d. Bottom: Eccentricity evolution for the injected Earth-mass
    planet at an initial semi-major axis of 1.735~AU.}
  \label{fig:sim}
\end{figure*}

Although the orbits of the four known giant planets within the
HD~141399 system are exceptionally stable (see Section~\ref{known}),
they may severely limit the possibility of stable terrestrial planet
orbits within the HZ. As described in Section~\ref{methods}, we tested
such scenarios via a suite of planet injection simulations for an
Earth-mass planet in the semi-major axis range 0.5--3.0~AU. The
percentage survival rates at each initial injected location are
represented in the top panel of Figure~\ref{fig:sim}. The extent of
the HZ is depicted using the same color scheme as for
Figure~\ref{fig:hz}, where light green and dark green indicate the CHZ
and OHZ, respectively.

As noted in Section~\ref{orbits}, planets c and d are the most massive
of the four known giant planets, each having masses $\sim$30\% larger
than Jupiter, and lie at either side of the HZ. As such, the
gravitational influence of these planets impose significant
constraints on the possible location of an injected Earth-mass planet
on either side of their orbits, indicated by vertical dashed lines in
Figure~\ref{fig:sim}. Many of the unstable regions are the result of
the MMR locations shown in Table~\ref{tab:mmr}, such as the 1:2 MMR
with planet c at 1.10~AU, and the 2:1 MMR with planet d at
1.33~AU. Even so, there are many islands of stability (albeit narrow)
that are present within the HZ, where stable orbits in this context
are defined as having survived the full $10^7$~year simulation at that
initial semi-major axis (see Section~\ref{methods}). Note that these
regions of stability mostly lie within the inner half of the HZ, and
the outer half of the HZ is dominated by the proximity to planet
d. Further note that there is an island of stability centered around
the semi-major axis of planet d, allowing for the possibility of
Trojan planets \citep{paez2015a}, for which it is feasible to maintain
long-term orbital stability \citep{cresswell2009,schwarz2009a}.

A 100\% survival of the simulation for the injected terrestrial planet
does not guarantee long-term stability beyond the $10^7$ years. Given
the narrow nature of the stability islands, the long-term stability of
the injected planet is highly sensitive to changes in its semi-major
axis and/or orbital eccentricity. An example is provided in the bottom
panel of Figure~\ref{fig:sim}, which shows the first $10^4$~years of
orbital evolution for the injected planet when the starting location
is $a = 1.735$~AU, corresponding to a very narrow stability spike in
the top panel of Figure~\ref{fig:sim}. The eccentricity variations for
the injected planet are dominated by a superposition of the perturbing
effects of planet c and d (see Figure~\ref{fig:known}), resulting in
an eccentricity amplitude similar to the sum of the planet c and d
eccentricity amplitudes. The long-term viability of such a scenario is
resolved soon after the $10^7$ year integration, during which an
additional simulation revealed the chaotic degradation of the injected
planet's orbit and ejection from the system. Although many of the
remaining stability islands shown in the top panel of
Figure~\ref{fig:sim} may indeed retain long-term stability, the
results presented here demonstrate the uncertain nature of evaluating
potential additional orbits within the HZ.

%%%%%%%%%%%%%%%%%%%%%%%%%%%%%%%%%%%%%%%%%%%%%%%%%%%%%%%%%%%%%%%%%%%%

\section{Discussion}
\label{discussion}

As described in Section~\ref{intro}, given the relative rarity of
giant planets, the HD~141399 system of four giant planets is
exceptionally rare among known exoplanetary architectures. According
to the NASA Exoplanet Archive \citep{akeson2013}, HD~141399 is one of
only two known planetary systems with at least four planets that are
all more massive than Saturn. The other system is HR~8799, with four
wide-separation planets that were detected via direct imaging
\citep{marois2008b,marois2010}. Thus, the HD~141399 system is an
incredible opportunity to study the formation, dynamics, and evolution
of an unusual planetary architecture.

The architecture of the HD~141399 system is known only to the level of
completeness of the RV measurements, and additional, as yet
undetected, planets may exist within the system. We have now entered
the era of extreme precision RV (EPRV) exoplanet searches
\citep{fischer2016} with the goal of detecting and measuring masses
for terrestrial planets. Such efforts have been applied to known
exoplanetary systems, such as the recent use of EPRV techniques to
detect two additional planets in the rho CrB system
\citep{brewer2023}, a system first discovered by \citet{noyes1997}. In
this work, we explored the dynamical feasibility of an Earth-mass
planet in the HZ which extends within the range 0.974--2.309~AU (see
Section~\ref{orbits}). The semi-amplitude of the RV signature for such
a planet would be 8.7~cm/s and 5.6~cm/s at the inner and outer edges
of the HZ, respectively. These amplitudes, like our dynamical
simulations, assume a near edge-on inclination for the system. One
possible resolution of the inclination ambiguity may lie with an
astrometric detection of the outer planet. Through the coming years,
the combination of RV and astrometric data will be an increasingly
powerful method to characterize non-transiting RV systems
\citep{perryman2014c,brandt2021a,winn2022}. Using the orbital
parameters provided by Table~\ref{tab:planets}, and the stellar
distance of 37.05~pc, the maximum astrometric amplitude of planet e is
0.07~mas. HD~141399 is relatively bright ($V = 7.21$), resulting in
the predicted astrometric amplitude for planet e lying well above the
Gaia DR3 astrometric precision \citep{brown2021,lindegren2021a}.
Indeed, the Hipparcos-Gaia Catalog of Accelerations
\citep{brandt2021a} reveals a moderately significant acceleration of
the star from Gaia DR3 astrometry when compared against the constant
proper motion model. Although there is no guarantee that the other
planets in the system are similarly inclined, an astrometric
measurement of the inclination for planet e would greatly aid in
understanding the formation and evolution of this relatively rare
planetary architecture.

Giant planet formation processes that follow the prescription of
models such as the ''Grand Tack'' model for the solar system
\citep[e.g.,][]{walsh2011c,raymond2014a,nesvorny2018c} suggest
disk-dominated giant planet migration occurs prior to the formation of
terrestrial planets \citep[e.g.,][]{chambers2014c}. Indeed,
observation and theoretical evidence supports the proposition that
giant planets form and migrate quickly relative to terrestrial
planets, thus shaping the architecture of such systems
\citep{pascucci2006,raymond2009c,morbidelli2012a,clement2023a}. This
implies that, should terrestrial planets be present in the HZ of
HD~141399, they likely formed after the known giant planets, in
particular planets c and d either side of the HZ. Such a formation
scenario would have significant effects on not only the formation and
dynamics of HZ planets, but also their impact history and hydration
\citep{obrien2014a,raymond2017b,sanchez2018,bailey2022b}. Thus, it
remains uncertain as to the planetary habitability outcome for
terrestrial planets within the HD~141399, even if stable orbits can be
retained.

It is worth noting that, from the perspective of a terrestrial planet
located near the center of the HD~141399 HZ, planet d would provide a
spectacular sight. From that perspective, and assuming Jupiter's
radius and geometric albedo, planet d would exhibit an apparent visual
magnitude of -10 when at opposition, shining more than 5 magnitudes
brighter than the maximum brightness of Venus as seen from
Earth. Additionally, planet d would have an angular size of
$\sim$7~arcminutes, which is $\sim$8.5 times larger than Jupiter as
seen from Earth, and would easily seen as a disk with the naked eye.

%%%%%%%%%%%%%%%%%%%%%%%%%%%%%%%%%%%%%%%%%%%%%%%%%%%%%%%%%%%%%%%%%%%%

\section{Conclusions}
\label{conclusions}

As the dominant planetary masses, and considering their relatively
rapid formation, giant planets play a pivotal role in the eventual
architecture and evolution of systems in which they are present. This
is especially important for systems that also contain terrestrial
planets, as the habitability of such planets will undoubtedly be
influenced by the presence of giant planets within the system.
HD~141399 stands as a particularly unusual system with respect to its
giant planet inventory, to the point where the pathway to the observed
architecture may have effectively suppressed terrestrial planet
formation within the system. If indeed there are terrestrial planets
that are present within the system HZ, they would serve as useful case
studies regarding the influence of giant planets on the prevalence and
evolution of surface habitability, such as the role of giant planets
in volatile delivery.

The HD~141399 dynamical study presented here provides the results for
two main investigations: the stability of the known planets and the
viability of a terrestrial planet within the HZ. The results show that
the known planets are long-term stable, and that the apsidal
trajectories for the b--c planet pair are librating, indicating that
they are in or near the 2:1 MMR. Our results for the Earth-mass planet
injection simulations show that there are limited islands of
stability, and that these islands are sensitive to the initial
conditions of the terrestrial planet. The combined angular momentum
transfers of planets c and d result in significant eccentricity
variations for the injected planet that can result in an eventual
chaotic orbit and loss from the system in some cases. With the
progress of RV and astrometric precision, it is hoped that key
systems, such as this one, will be observationally revisited in the
years ahead to better understand the nature of giant planet systems
and their potential to harbor habitable planets in their midst.

%%%%%%%%%%%%%%%%%%%%%%%%%%%%%%%%%%%%%%%%%%%%%%%%%%%%%%%%%%%%%%%%%%%%

\section*{Acknowledgements}

This research has made use of the Habitable Zone Gallery at
hzgallery.org. The results reported herein benefited from
collaborations and/or information exchange within NASA's Nexus for
Exoplanet System Science (NExSS) research coordination network
sponsored by NASA's Science Mission Directorate.

%%%%%%%%%%%%%%%%%%%%%%%%%%%%%%%%%%%%%%%%%%%%%%%%%%%%%%%%%%%%%%%%%%%%

\software{Mercury \citep{chambers1999}}

%%%%%%%%%%%%%%%%%%%%%%%%%%%%%%%%%%%%%%%%%%%%%%%%%%%%%%%%%%%%%%%%%%%%

%\bibliographystyle{aasjournal}
%\bibliography{/data/skane/latex/styles/references}

%%%%%%%%%%%%%%%%%%%%%%%%%%%%%%%%%%%%%%%%%%%%%%%%%%%%%%%%%%%%%%%%%%%%

\end{document}